# Capacity Theorems for the AWGN Multi-Way Relay Channel

Lawrence Ong, Christopher M. Kellett, and Sarah J. Johnson

## Abstract

The $L$-user additive white Gaussian noise multi-way relay channel is considered, where multiple users exchange information through a single relay at a common rate. Existing coding strategies, i.e., complete-decode-forward and compress-forward are shown to be bounded away from the cut-set upper bound at high signal-to-noise ratios (SNR). It is known that the gap between the compress-forward rate and the capacity upper bound is a constant at high SNR, and that between the complete-decode-forward rate and the upper bound increases with SNR at high SNR. In this paper, a functional-decode-forward coding strategy is proposed. It is shown that for $L \geq 3$, complete-decode-forward achieves the capacity when SNR $\leq 0$ dB, and functional-decode-forward achieves the capacity when SNR $\geq 0$ dB. For $L = 2$, functional-decode-forward achieves the capacity asymptotically as SNR increases.

## 1 Introduction

In this paper, we consider the additive white Gaussian noise (AWGN) multi-way relay channel (MWRC), in which $L$ users exchange full information at a common rate via a relay. When all nodes are subject to the same power constraint, we find:

- for $L \geq 3$, the capacity,
- for $L = 2$, asymptotic capacity results as SNR increases.

It has been shown that the complete-decode-forward[1] coding strategy performs poorly at high SNR, and the compress-forward coding strategy achieves rates within a constant number of bits of the capacity for all SNR [1]. In this paper, we show that complete-decode-forward achieves the capacity for SNR $\leq 0$ dB when there are more than two users. However, there is still a finite gap between the achievable rates and the cut-set upper bound at medium to high SNR.

---

This work is supported by the Australian Research Council under grant DP0877258.

[1] We modified the strategy name "decode-and-forward" used in the original paper [1] to distinguish this coding strategy and our proposed functional-decode-forward coding strategy.

We use a functional-decode-forward coding strategy, where the relay decodes functions of the users' messages and broadcasts the functions back to the users. The functions are defined such that combining the functions and its own message, every user is able to decode the messages of all other users. We close the gap between the capacity upper bound and achievable rates by showing that functional-decode-forward achieves the capacity for SNR $\geq 0$ dB when there are more than two users. For two users, functional-decode-forward achieves the capacity asymptotically as SNR increases.

## 2 Preliminary

We first introduce the concept of functional-decode-forward for the MWRC by using a simple three-user example. In this paper, we denote by $X_i$ node $i$'s input to the channel, $Y_i$ the channel output received by node $i$, and $W_i$ node $i$'s message.

The idea of functional-decode-forward is for the relay to decode only functions of the users' messages, and the functions, when broadcast from the relay back to the users, are *merely* sufficient for them to decode other users' messages. The more information the relay needs to decode, the lower the rates the users can send on the *uplink*. For the two-user MWRC, the modular sum of the users' codewords is a good choice of function for certain types of channels, e.g., the AWGN two-way relay channel [2, 3]. However, decoding the modular sum of all users' codewords will not work for MWRCs with more than two users. Thus, for the general $L$-user case, the best function for the relay to decode is not obvious.

In this paper, we propose that the relay decodes and forwards *functions of message pairs*. To illustrate this strategy, we consider the following three-user noiseless MWRC:

- uplink: $Y_0 = X_1 \oplus X_2 \oplus X_3$,
- downlink: $Y_1 = Y_2 = Y_3 = X_0$,

where $X_i \in \{0, 1\}$, $\forall i$, and $\oplus$ is modulo-two addition. We assume that the messages $W_i$, $\forall i$, are random bits. The users split their transmissions into two phases. In the first phase, the users send $X_1 = W_1$, $X_2 = W_2$, $X_3 = 0$. In the second phase, the users send $W_1 = 0$, $X_2 = W_2$, $X_3 = W_3$. At the end of the two phases, the relay has $(W_1 \oplus W_2)$ and $(W_2 \oplus W_3)$. It then broadcasts these combined messages back to the users in two channel uses. After getting the messages from the relay, user 1 can obtain $W_2 = (W_1 \oplus W_2) \oplus W_1$ followed by $W_3 = (W_2 \oplus W_3) \oplus W_2$. User 2 can recover $(W_1, W_3)$ and user 3 can recover $(W_1, W_2)$ similarly.

Using this functional-decode-forward strategy, each user can send 1 message bit in two channel uses, i.e., each user can transmit at the rate of $\frac{1}{2}$ bit/channel use. On the downlink, each user can receive a maximum of 1 information bit/received symbol. Since each user must decode two users' messages (of 1 bit each), the capacity of this MWRC cannot exceed $\frac{1}{2}$ bit/channel use. So, the functional-decode-forward achieves the capacity in this example. Note that the capacity-achieving functional-decode-forward coding strategy is not unique.

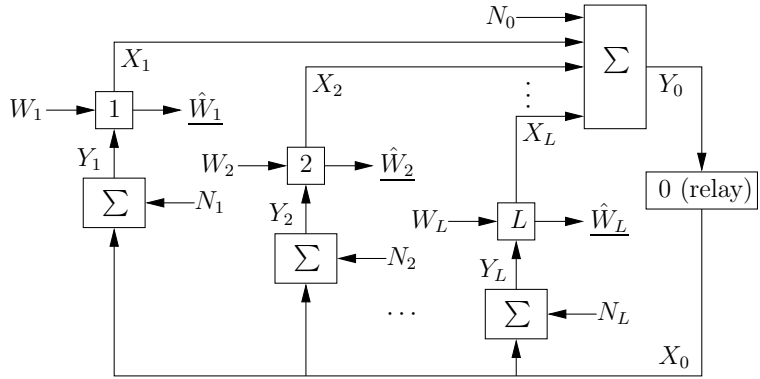

Figure 1: An $L$-user AWGN MWRC, where $\hat{\underline{W}}_i$ is user $i$'s estimate of all other users' messages

Had the relay used the complete-decode-forward coding strategy, it would need at least three channel uses to decode all three messages in the uplink. Hence the maximum rate achievable for each message is $\frac{1}{3}$ bit/channel use.

## 3 Channel Model

Fig. 1 depicts the $L$-user AWGN MWRC considered in this paper, where the uplink and the downlink channels are separated, i.e., there is no direct user-to-user link. Nodes 1 to $L$ are the users, and node 0 is the relay. By definition, $L \geq 2$. Each user is to decode the messages from all other users.

**Definition 1.** *We define the AWGN MWRC as follows:*

- *The uplink channel is the sum of all users' channel inputs and the relay's receiver noise:*

$$Y_0 = \sum_{i=1}^{L} X_i + N_0, \tag{1}$$

  *where the $X_i$ are subject to the power constraints $E[X_i^2] \leq P_i$, and $N_0$ is an i.i.d. zero-mean Gaussian random variable with unit variance $E[N_0^2] = \sigma_0^2 = 1$.*

- *The downlink consists of an independent point-to-point AWGN channel for each user, $i = 1, 2, \ldots, L$, i.e.,*

$$Y_i = X_0 + N_i, \tag{2}$$

  *where $X_0$ is subject to the power constraint $E[X_0^2] \leq P_0$, and $N_i$ is user $i$'s receiver noise and is an i.i.d. zero-mean Gaussian random variable with unit variance $E[N_i^2] = \sigma_i^2 = 1$.*

We consider the restricted MWRC in the sense that the transmit signals of each user can only depend on its message, and cannot depend on its received signals. We consider the following block code of $n$ channel uses:

**Definition 2.** *A $(2^{nR_1}, 2^{nR_2}, \ldots, 2^{nR_L}, n)$ code for the MWRC consists of*

1. *$L$ sets of messages: $W_i \in \mathcal{W}_i = \{1, 2, \ldots, 2^{nR_i}\}$, $i = 1, 2, \ldots, L$.*

2. *$L$ user encoding functions: $\boldsymbol{X}_i(W_i) = f_i(W_i)$, $i = 1, 2, \ldots, L$.*

3. *A set of relay encoding functions: $X_0[t] = f_{0,t}(Y_0[1], Y_0[2], \ldots, Y_0[t-1])$, $t = 1, \ldots, n$.*

4. *$L$ user decoding functions: $\underline{\hat{W}_i} \triangleq (\hat{W}_{i,1}, \ldots, \hat{W}_{i,i-1}, \hat{W}_{i,i+1}, \ldots, \hat{W}_{i,L}) = g_i(\boldsymbol{Y}_i, W_i)$, $i = 1, \ldots, L$, where $\hat{W}_{i,j}$ is node $i$'s estimate of $W_j$.*

In this paper, bold letters are used to define vectors of length $n$, e.g., $\boldsymbol{X} = (X[1], X[2], \ldots, X[n])$.

**Definition 3.** *Assuming that the message tuple $\omega \triangleq (W_1, W_2, \ldots, W_L)$ is uniformly distributed over the product set $\Omega \triangleq \mathcal{W}_1 \times \mathcal{W}_2 \times \cdots \times \mathcal{W}_L$, the* average error probability *for the $(2^{nR_1}, 2^{nR_2}, \ldots, 2^{nR_L}, n)$ code is defined as*

$$P_e = \Pr\left\{\hat{W}_{i,j} \neq W_j, \text{for some} j \in [1, L] \text{and some} i \neq j\right\}$$

$$= \frac{1}{2^{n\sum_{j=1}^{L} R_j}} \sum_{\alpha \in \Omega} \Pr\left\{\bigcup_{1 \leq i \leq L} g_i(\boldsymbol{Y}_i, W_i) \neq \alpha_{-i} \middle| \omega = \alpha\right\},$$

*where $\alpha_{-i}$ is defined as $\alpha$ without the $i$-th entry.*

**Definition 4.** *A rate tuple $(R_1, R_2, \ldots, R_L)$ is said to be* achievable *if, for any $\epsilon > 0$, there is at least one $(2^{nR_1}, 2^{nR_2}, \ldots, 2^{nR_L}, n)$ code such that $P_e < \epsilon$.*

We say that a node can *reliably* decode a message if and only if the average probability that the node wrongly decodes the message can be made arbitrarily small.

In this paper, we assume that $P_i = P$, $\forall i \in [1, L]$, and we focus on the *common rate* $R = R_i$, $\forall i \in [1, L]$. We say that the common rate $R$ is achievable if the rate tuple $(R, R, \ldots, R)$ is achievable.

**Definition 5.** *We define the* common-rate capacity *of the MWRC as (also known as the symmetrical capacity [1])*

$$C_{\mathrm{CR}} \triangleq \sup\{R : (R, R, \ldots, R) \text{is achievable}\}. \tag{4}$$

The common rate is useful in systems where all users have the same amount of information to send, or in *fair* systems where every user is to be given the same guaranteed uplink *bandwidth*, i.e., each user can send data up to a certain rate.

# 4 Existing Results

## 4.1 Capacity Upper Bound

An upper bound to the common-rate capacity of the AWGN MWRC based on cut-set arguments (see [4, page 589 (Theorem 15.10.1)]) is given by:

**Proposition 1** ( [1, Proposition 1] with one cluster). *The common-rate capacity of the AWGN MWRC is upper-bounded by*

$$C_{\text{CR}} \leq \min\left\{\frac{1}{2(L-1)}\log[1+(L-1)P], \frac{1}{2(L-1)}\log[1+P_0]\right\} \triangleq R_{\text{UB}}. \quad (5)$$

In this paper, log denotes logarithm to the base two, and hence the rates are in bits/channel use.

## 4.2 Complete-Decode-Forward

Using the complete-decode-forward coding strategy, the relay decodes all users' messages, encodes and broadcasts a function of the messages back to the users. We have

**Proposition 2** ( [1, Proposition 3] with one cluster). *Consider an L-user AWGN MWRC. Complete-decode-forward achieves the following common rate:*

$$R_{\text{CDF}} = \min\left\{\frac{1}{2L}\log[1+LP], \frac{1}{2(L-1)}\log[1+P_0]\right\}. \quad (6)$$

## 4.3 Compress-Forward

Using the compress-forward coding strategy, the relay quantizes its received signals, encodes and broadcasts them to the users. We have

**Proposition 3** ( [1, Proposition 4] with one cluster). *Consider an L-user AWGN MWRC. Compress-forward achieves the following common rate:*

$$R_{\text{CF}} = \frac{1}{2(L-1)}\log\left[1 + \frac{(L-1)PP_0}{1+(L-1)P+P_0}\right]. \quad (7)$$

**Remark 1.** *It has been shown [1, Remark 2] that the compress-forward coding strategy always achieves a higher common rate than the amplify-forward coding strategy does.*

# 5 Functional-Decode-Forward Coding Strategy

Our proposed functional-decode-forward coding strategy for the AWGN MWRC is based on lattice codes. We first review some basics of lattice codes. An $n$-dimensional lattice $\Lambda$ is a discrete subgroup of the $n$-dimensional Euclidean space $\mathfrak{R}^n$ under the normal vector addition operation. This means if $\boldsymbol{v}_1, \boldsymbol{v}_2 \in \Lambda$, then $v_1 + v_2 \in \Lambda$. For any $x \in \Re^n$, a modulo-$\Lambda$ operation is is defined as: $x \mod \Lambda = x - Q_\Lambda(x)$, where $Q_\Lambda(x) \in \Lambda$ is the lattice point that is closest to $x$. The fundamental Voronoi region $\mathcal{V}(\Lambda)$ for a lattice $\Lambda$ is the set of all points in $\Re^n$ that are closer to the origin than they are to any other lattice point, i.e., $\mathcal{V}(\Lambda) = \{x \in \Re^n : Q_\Lambda(x) = 0\}$, where $0$, the origin, is the all-zero vector of length $n$.

For lattice encoding, we consider two lattices, where the coarse lattice $\Lambda$ is nested in the fine lattice $\Lambda_f$, i.e., $\Lambda \subseteq \Lambda_f$. A message $w$ is mapped to a fine lattice point that sits in the fundamental Voronoi region of the course lattice $\Lambda$, i.e., $v(w) \in \{\Lambda_f \cap \mathcal{V}(\Lambda)\}$. $\Lambda$ is selected such that the the transmit power constraint of all the users can be met, and $\Lambda_f$ is selected such that there are $2^{nR}$ fine lattice points in $\{\Lambda_f \cap \mathcal{V}(\Lambda)\}$.

## 5.1 Uplink

The uplink transmissions are split into $(L-1)$ blocks of $n$ channel uses each. In block $l$, $1 \leq l \leq L-1$, nodes $l$ and $(l+1)$ transmit using lattice codes, and all other nodes do not transmit, i.e.,

$$\boldsymbol{X}_i(W_i) = \begin{cases} \boldsymbol{V}(W_i) + \boldsymbol{d}_i \mod \Lambda, & \text{if } i = l, l+1 \\ \boldsymbol{0}, & \text{otherwise,} \end{cases} \tag{8}$$

where $\boldsymbol{0}$ is the all-zero vector of length $n$, $\boldsymbol{V}(W_i) \in \{\Lambda_f \cap \mathcal{V}(\Lambda)\}$ contains user's information $W_i$, and $\boldsymbol{d}_i \in \Re^n$ is an independently and randomly generated vector uniformly distributed over $\mathcal{V}(\Lambda)$ which is fixed for all transmissions.

As the codewords for all users are uniformly distributed in $\mathcal{V}(\Lambda)$, all users transmit at the same power, $P'$. For $L \geq 3$, since nodes 2 to $(L-1)$ only transmit in two of the $(L-1)$ blocks, and nodes 1 and $L$ transmit in one block, we can set all nodes to transmit at $P' = \frac{L-1}{2}P$ while still satisfying the average power constraint of $E[X_i^2] \leq P$. For $L = 2$, there is only one block, and both the users transmit at power $P' = P$.

In block $l$, the relay decodes $\boldsymbol{V}_{l,l+1} \triangleq \bigl(\boldsymbol{V}(W_l) + \boldsymbol{V}(W_{l+1})\bigr) \mod \Lambda$, which is a function of the messages $W_l$ and $W_{l+1}$. Doing this for all $(L-1)$ blocks, the relay can reliably decode the functions $(\boldsymbol{V}_{1,2}, \boldsymbol{V}_{2,3}, \ldots, \boldsymbol{V}_{L-1,L})$ if [2,3]

$$R \leq \frac{1}{L-1}\left\{\frac{1}{2}\log\left[\frac{1}{2} + P'\right]\right\}^+, \tag{9}$$

with a sufficiently large $n$. Here, $P'$ is the transmit power of each user, given by

$$P' = \begin{cases} P, & \text{if } L = 2 \\ \frac{L-1}{2}P, & \text{otherwise}(L \geq 3), \end{cases} \tag{10}$$

and $x^+ = \max\{x, 0\}$. The factor $\frac{1}{L-1}$ in (9) takes into account that the transmission and decoding for each $\boldsymbol{V}_{l,l+1}$ only happens in one of the $(L-1)$ blocks.

## 5.2 Downlink

Now, since $\boldsymbol{V}_{l,l+1} \in \{\Lambda_f \cap \mathcal{V}(\Lambda)\}$, $\forall l \in [1, L-1]$, there are at most $2^{n(L-1)R}$ unique vectors $(\boldsymbol{V}_{1,2}, \boldsymbol{V}_{2,3}, \ldots, \boldsymbol{V}_{L-1,L})$. In the downlink, the relay broadcasts this vector back to all the users. As the downlink to each user is a point-to-point AWGN channel, each user can reliably decode $(\boldsymbol{V}_{1,2}, \boldsymbol{V}_{2,3}, \ldots, \boldsymbol{V}_{L-1,L})$ if

$$(L-1)R \leq \frac{1}{2}\log[1+P_0], \tag{11}$$

with a sufficiently large $n$.

Assuming that user $i$, $i \in [1, L]$, is able to correctly decode $(\boldsymbol{V}_{1,2}, \boldsymbol{V}_{2,3}, \ldots, \boldsymbol{V}_{L-1,L})$ sent by the relay, it performs the following (the order of decoding is important) to obtain all other users' messages:

$$\boldsymbol{V}(W_{i+1}) = (\boldsymbol{V}_{i,i+1} - \boldsymbol{V}(W_i)) \mod \Lambda \tag{12}$$
$$\boldsymbol{V}(W_{i+2}) = (\boldsymbol{V}_{i+1,i+2} - \boldsymbol{V}(W_{i+1})) \mod \Lambda \tag{13}$$
$$\cdots$$
$$\boldsymbol{V}(W_L) = (\boldsymbol{V}_{L-1,L} - \boldsymbol{V}(W_{L-1})) \mod \Lambda \tag{14}$$
$$\boldsymbol{V}(W_{i-1}) = (\boldsymbol{V}_{i-1,i} - \boldsymbol{V}(W_i)) \mod \Lambda \tag{15}$$
$$\boldsymbol{V}(W_{i-2}) = (\boldsymbol{V}_{i-2,i-1} - \boldsymbol{V}(W_{i-1})) \mod \Lambda \tag{16}$$
$$\cdots$$
$$\boldsymbol{V}(W_1) = (\boldsymbol{V}_{1,2} - \boldsymbol{V}(W_2)) \mod \Lambda. \tag{17}$$

## 5.3 Achievability

Combining the uplink and the downlink, if (9) and (11) are satisfied, all users can reliably decode the messages of all other users. So, using this functional-decode-forward coding strategy, the following common rate is achievable:

$$R = \min\left\{\left\{\frac{\log\left[\frac{1}{2} + P'\right]}{2(L-1)}\right\}^+, \frac{\log[1+P_0]}{2(L-1)}\right\}, \tag{18}$$

where $P'$ is defined in (10).

**Remark 2.** *The strategy proposed here is different from the strategy described in [1, Section IV.B.] (also using lattice codes), where there are more than one cluster with two users in each cluster, and only the two users in each cluster exchange messages. In the MWRC considered in this paper, there is only one cluster with L users and all users engage in full data exchange.*

## 5.4 An Improved Functional-Decode-Forward

Although we have shown that the functional-decode-forward coding strategy described in Sections 5.1 and 5.2 achieves the capacity of the binary MWRC [5], it is not always optimal in the AWGN counterpart. Now, we slightly modify this strategy to improve its rate in the AWGN MWRC, i.e., (18). We have seen that on the uplink for $L \geq 3$, nodes 1 and $L$ only transmit in one of the $(L-1)$ transmission blocks, while the other nodes transmit in two of the $(L-1)$ transmission blocks. Setting $P' = \frac{L-1}{2}P$, nodes 1 and $L$ are not transmitting at their maximum allowable power. Consider multiple messages for each user, and let the $t$-th message tuple be $(W_1[t], W_2[t], \ldots, W_L[t])$, for $t \in \{1, 2, 3, \ldots\}$. For each $t$, we have $(L-1)$ blocks of transmissions. Instead of fixing nodes $l$ and $(l+1)$ to transmit in block $l$, $1 \leq l \leq L-1$, for all message tuples, we *rotate* the transmission scheme for each message tuple such that in block $l$, nodes $(l+t-2 \mod L)+1$ and $(l+t-1 \mod L)+1$ transmit, and all the other nodes do not transmit, i.e., $\boldsymbol{X}_{i+1}(W_{i+1}[t]) =$

$$\begin{cases} \boldsymbol{V}(W_{i+1}[t]) + \boldsymbol{d}_{i+1} \mod \Lambda, & \text{if } i = l+t-2 \mod L, \\ & \text{or } i = l+t-1 \mod L \\ \boldsymbol{0}, & \text{otherwise.} \end{cases}$$

The above transmission scheme repeats itself after every $L$ message tuples. Consider a window of $L$ message tuples, e.g., $t \in [1, L]$. As there are $(L-1)$ blocks of transmissions for each message tuple, there are all together $L(L-1)$ blocks of transmissions. Each node transmits in only one block for two of the $L$ message tuples, and transmits in two blocks for the other $(L-2)$ message tuples. So, each node can transmit with $\frac{L(L-1)}{2+2(L-2)}P$, giving an average transmit power of $E[X_i^2] = P$. So, for this *improved* functional-decode-forward coding strategy, the transmit power of each node, in (10), can be increased to $P' = \frac{L(L-1)}{2+2(L-2)} = \frac{L}{2}P$. Note that this is also true for $L = 2$ where both users transmit all the time. Also, note that under this transmission scheme, when the relay broadcasts all $\{\boldsymbol{V}_{i,j}\}$ back to the users, each user can decode other users' messages using the method described in Section 5.2. This gives the following achievable rate:

**Theorem 1.** *Consider an L-user AWGN MWRC. Functional-decode-forward achieves the following common rate:*

$$R_{\text{FDF}} = \min\left\{\left\{\frac{1}{2(L-1)} \log\left[\frac{1}{2} + \frac{L}{2}P\right]\right\}^+, \frac{1}{2(L-1)} \log[1+P_0]\right\}. \quad (19)$$

## 6 The Common-Rate Capacity of the AWGN MWRC

In this section, we consider the case where the transmit power of all users and the relay is equal, i.e., $P_0 = P$. This means $\text{SNR} = \frac{P}{\sigma_0^2} = P$ for the relay, and $\text{SNR} = \frac{P_0}{\sigma_i^2} = P$ for every user $i$ (recall that $\sigma_j^2 = 1$, $\forall j \in [0, L]$, by definition).

## 6.1 Upper Bound

When $P_0 = P$, the upper bound on the common-rate capacity in Proposition 1 simplifies to

$$R_{\text{UB}} = \frac{1}{2(L-1)} \log[1+P]. \quad (20)$$

## 6.2 Functional-Decode-Forward

First, we show that functional-decode-forward achieves the common-rate capacity under certain conditions.

**Theorem 2.** *Consider the AWGN MWRC with $P_0 = P$.*

- *For $L \geq 3$: if $P \geq \frac{1}{L-2}$, the common-rate capacity is*

$$C_{\text{CR}} = \frac{1}{2(L-1)} \log[1+P], \quad (21)$$

  *and it is achievable by functional-decode-forward.*

- *For $L = 2$: the common-rate capacity is bounded by*

$$\left\{\frac{\log\left[\frac{1}{2}+P\right]}{2}\right\}^+ \leq C_{\text{CR}} \leq \frac{\log[1+P]}{2} < \left\{\frac{\log\left[\frac{1}{2}+P\right]}{2}\right\}^+ + \epsilon(P), \quad (22)$$

  *where $\epsilon(P) = \min\left\{\frac{1}{2}, \frac{1}{2(2P+1)\ln 2}\right\} \xrightarrow{P \to \infty} 0$.*
  *Functional-decode-forward achieves rates within $\frac{1}{2}$ bit of the capacity, and achieves the common-rate capacity asymptotically as $P$ increases.*

**Remark 3.** *For $L = 2$, if we consider the gap $\epsilon(P)$ normalized to the capacity upper bound, we have*

$$\frac{\epsilon(P)}{R_{\text{UB}}} \leq \frac{1}{(2P+1)\ln[1+P]}. \quad (23)$$

*So, functional-decode-forward achieves the common-rate capacity asymptotically as $P$ increases in an absolute sense as well as in a normalized (to the upper bound) sense.*

*Proof of Theorem 2.* For $L \geq 3$, if $P \geq \frac{1}{L-2}$, we have

$$1 + LP \geq 2 + 2P$$

$$\frac{1}{2(L-1)} \log\left[\frac{1}{2} + \frac{L}{2}P\right] \geq \frac{1}{2(L-1)} \log[1+P].$$

So, from (19) and (20), $R_{\text{FDF}} = \frac{1}{2(L-1)} \log[1+P] = R_{\text{UB}}$.

Next, for $L = 2$, we have $R_{\text{FDF}} = \frac{1}{2} \log\left[\frac{1}{2} + P\right]$ as the first term is smaller than the second term on the RHS of (19). Note that $\frac{d}{dx} \log[x] = \frac{1}{x \ln 2}$ and $\frac{d^2}{dx^2} \log[x] = -\frac{1}{x^2 \ln 2} < 0$. So,

$$\log[x + \delta] < \log[x] + \left.\frac{d}{dy} \log[y]\right|_{y=x} ((x + \delta) - x)$$
$$= \log[x] + \frac{\delta}{x \ln 2}.$$

Hence, from (20),

$$R_{\text{UB}} = \frac{1}{2} \log[1 + P]$$
$$< \frac{1}{2} \log\left[\frac{1}{2} + P\right] + \frac{1}{2} \frac{\frac{1}{2}}{\left(P + \frac{1}{2}\right) \ln 2}$$
$$= R_{\text{FDF}} + \frac{1}{2(2P + 1)(L - 1) \ln 2}.$$

Furthermore, $R_{\text{UB}} = \frac{1}{2} \log\left[2\left(\frac{1}{2} + \frac{P}{2}\right)\right] < \frac{1}{2} \log\left[\frac{1}{2} + P\right] + \frac{1}{2} \log 2 = R_{\text{FDF}} + \frac{1}{2}$. Since, $R_{\text{FDF}} \leq C_{\text{CR}} \leq R_{\text{UB}}$, we have Theorem 2. □

### 6.3 Complete-Decode-Forward

From Proposition 2, we have

$$R_{\text{CDF}} = \min\left\{\frac{1}{2L} \log[1 + LP], \frac{1}{2(L-1)} \log[1 + P]\right\}. \quad (27)$$

We first derive the region in which complete-decode-forward achieves the common-rate capacity.

**Theorem 3.** *Consider the AWGN MWRC with $P_0 = P$.*

- *For $L \geq 3$: if $0 < P \leq 1$, the common-rate capacity is*

$$C_{\text{CR}} = \frac{1}{2(L-1)} \log[1 + P], \quad (28)$$

 *and it is achievable by complete-decode-forward.*

- *For $L = 2$: $R_{\text{CDF}} < R_{\text{UB}}$, i.e., the complete-decode-forward rate is strictly below the capacity upper bound.*

*Proof of Theorem 3.* Define $\alpha(L, P) = \left(\frac{1+LP}{1+P}\right)^{L-1}$ and $\beta(P) = 1 + P$. From (20) and (27), we can show that $R_{\text{CDF}} = R_{\text{UB}}$ iff $\alpha(L, P) \geq \beta(P)$, and $R_{\text{CDF}} < R_{\text{UB}}$ otherwise. Note that $\alpha(L, 0) = \beta(0)$, and $\frac{d}{dP} \beta(P) = 1$ for $\forall P$. In addition,

$$\frac{d}{dP} \alpha(L, P) = (L - 1)^2 (1 + P)^{-L} (1 + LP)^{L-2} > 0$$
$$\frac{d^2}{dP^2} \alpha(L, P) = \frac{(L - 1)^2 L (1 + LP)^{L-3} (L - 3 - 2P)}{(1 + P)^{L+1}}.$$

For $L = 2$, $\frac{d}{dP}\alpha(L,P)\big|_{P=0} = 1$ and $\frac{d^2}{dP^2}\alpha(L,P) < 0$. So $\alpha(L,P) < \beta(P)$, and $R_{\text{CDF}} < R_{\text{UB}}$, $\forall P > 0$.

For $L \geq 3$, $\frac{d}{dP}\alpha(L,P)\big|_{P=0} > 1$, $\frac{d^2}{dP^2}\alpha(L,P)$ decreases as $P$ increases, and $\frac{d^2}{dP^2}\alpha(L,P) < 0$ when $P > \frac{L-3}{2}$. So, there exists a point $P^*(L) > 0$, where $\alpha(L,P) \geq \beta(P)$ for $P \leq P^*(L)$, and $\alpha(L,P) < \beta(P)$ for $P > P^*(L)$. If we fix $P = 1$, since $L \geq 3$, we have $\frac{1+L}{2} \geq 2$, meaning that $\left(\frac{1+LP}{1+P}\right)^{L-1} \geq 1 + P$, and therefore $\alpha(L,P) \geq \beta(P)$. So, $P = 1$ falls into the region in which $\alpha(L,P) \geq \beta(P)$. This gives $P^*(L) \geq 1$. Hence, for $L \geq 3$ and $0 < P \leq 1$, we have $R_{\text{CDF}} = R_{\text{UB}}$. □

Next, we show that the complete-decode-forward rate is bounded below the capacity upper bound at large $P$.

**Theorem 4.** *Consider the AWGN MWRC with $P_0 = P$. If $P > L^{L-1} - 1$, then $R_{\text{CDF}} < R_{\text{UB}}$. Furthermore, for any finite $L$, as $P \to \infty$, the maximum complete-decode-forward rate is $\left(\frac{\log[1+P] - (L-1)\log L}{2L(L-1)}\right)$ bits below the common-rate capacity.*

The gap between $R_{\text{CDF}}$ and $R_{\text{UB}}$ increases with $P$ as $P \to \infty$.

*Proof of Theorem 4.* From (27), if $\frac{1}{2L}\log[1 + LP] \leq \frac{1}{2(L-1)}\log[1 + P]$, then $R_{\text{CDF}} = \frac{1}{2L}\log[1 + LP]$. Now,

$$\frac{1}{2(L-1)}\log[1+P] - \frac{1}{2L}\log[1+LP]$$
$$\stackrel{(\varphi)}{=} \frac{1}{2L(L-1)}\left(\log[1+P] - (L-1)\log\left[\frac{LP+1}{P+1}\right]\right)$$
$$= \frac{1}{2L(L-1)}\left(\log[1+P] - (L-1)\log\left[\frac{L+\frac{1}{P}}{1+\frac{1}{P}}\right]\right).$$

Since $L \geq 2$ and $P > 0$, we have $1 \leq \left(\frac{LP+1}{P+1}\right)^{L-1} \leq L^{L-1}$. So, if $P > L^{L-1} - 1$, the RHS of $(\varphi)$ is strictly positive. Under this condition, $R_{\text{CDF}} = \frac{1}{2L}\log[1 + LP] < R_{UB} = \frac{1}{2(L-1)}\log[1+P]$. From Theorem 2, $\lim_{P \to \infty} R_{\text{UB}} = C_{\text{CR}}$. □

## 6.4 Compress-Forward

Now, we show that the compress-forward rate is bounded below the common-rate capacity upper bound at all $P$.

**Theorem 5.** *Consider the AWGN MWRC with $P_0 = P$. Compress-forward achieves rates up to $\left(\frac{1}{2(L-1)}\log\left[1 + \frac{P}{(L-1)P+1}\right]\right)$ bits below the capacity upper bound for all $L$ and $P$. Furthermore, for any finite $L$, as $P \to \infty$, the maximum compress-forward rate is $\left(\frac{1}{2(L-1)}\log\frac{L}{L-1}\right)$ bits below the common-rate capacity.*

**Remark 4.** *Here, the gap $\left(\frac{1}{2(L-1)} \log\left[1 + \frac{(L-1)P}{1+LP}\right]\right)$ is strictly smaller than $\frac{1}{2(L-1)}$ stated in [1, Theorem 1].*

*Proof of Theorem 5.* From Proposition 3,

$$R_{\text{CF}} = \frac{1}{2(L-1)} \log\left[1 + \frac{(L-1)P^2}{1 + (L-1)P + P}\right]$$
$$= \frac{1}{2(L-1)} \log[1+P] + \frac{1}{2(L-1)} \log\left[\frac{1 + (L-1)P}{1 + LP}\right]$$
$$= R_{\text{UB}} - \frac{1}{2(L-1)} \log\left[1 + \frac{1}{(L-1) + \frac{1}{P}}\right].$$

From Theorem 2, $\lim_{P \to \infty} R_{\text{UB}} = C_{\text{CR}}$. □

## 6.5 The Common-Rate Capacity

Combining Theorems 2 and 3, we have the following capacity results:

**Theorem 6.** *Consider the AWGN MWRC with $P_0 = P$.*

- *For $L \geq 3$: The common-rate capacity is $C_{\text{CR}} = \frac{\log[1+P]}{2(L-1)}$, and is achievable by*
  - *complete-decode-forward, if $0 < P \leq 1$, and*
  - *functional-decode-forward, otherwise, i.e., $P > 1$.*

- *For $L = 2$: $\left\{\frac{\log\left[\frac{1}{2}+P\right]}{2}\right\}^+ \leq C_{\text{CR}} \leq \frac{\log[1+P]}{2} < \left\{\frac{\log\left[\frac{1}{2}+P\right]}{2}\right\}^+ + \epsilon(P)$, where $\lim_{P \to \infty} \epsilon(P) = 0$.*
  - *Functional-decode-forward (where $R_{\text{FDF}} = \left\{\frac{1}{2}\log\left[\frac{1}{2} + P\right]\right\}^+$) achieves the common-rate capacity asymptotically as $P$ increases.*